\documentclass{article}

\usepackage{arxiv}

\usepackage[utf8]{inputenc} 
\usepackage[T1]{fontenc}    
\usepackage{hyperref}       
\usepackage{url}            
\usepackage{booktabs}       
\usepackage{amsfonts}       
\usepackage{nicefrac}       
\usepackage{microtype}      
\usepackage{lipsum}		
\usepackage{graphicx}
\usepackage{amsmath}
\usepackage{subcaption}

\title{Geodesic Loewner paths with varying boundary conditions}

\date{June  2020}	

\author{ N.R. McDonald \\
	Department of Mathematics\\
	University College London\\
	London, WC1E 6BT \\
	\texttt{n.r.mcdonald@ucl.ac.uk} \\
}

\renewcommand{\headeright}{}
\renewcommand{\undertitle}{}

\hypersetup{
pdftitle={Geodesic Loewner paths with varying boundary conditions},
pdfauthor={N.R. McDonald},
pdfkeywords={Loewner equation, Laplacian growth, free boundary problems},
}

\begin{document}
\maketitle

\begin{abstract}
Equations of the Loewner class subject to  non-constant boundary conditions along the real axis, are formulated and solved giving the geodesic paths of slits growing in the upper half  complex plane. The problem is motivated by Laplacian growth in which the slits represent thin fingers growing in a diffusion field. A single finger  follows a curved path  determined by the forcing function appearing in Loewner's equation. This function is found  by solving an ordinary differential equation whose terms depend on curvature properties of the streamlines of the diffusive field in the conformally mapped `mathematical' plane. The effect of boundary conditions specifying either piecewise constant values of the field variable along the real axis, or  a dipole placed on the real axis, reveal a range of   behaviours for the growing slit. These include regions along the real axis from which no slit growth is possible, regions where paths grow to infinity, or regions where paths curve back toward the real axis terminating in finite time. Symmetric pairs of paths subject to the piecewise constant boundary condition along the real axis are also computed, demonstrating that  paths which grow to infinity  evolve asymptotically toward an angle of bifurcation of $\pi/5$.	
\end{abstract}


\section{Introduction}
\label{intro}

The evolution of an interface separating two different phases frequently leads to intricate patterns  characterised by  finger-like intrusions of one phase penetrating the other e.g. \cite{saff,zik,giverso}.  When the phase dynamics is governed by Laplace's equation, and the interface velocity given by the local gradient of the phase, the evolution of the interface is referred to as Laplacian growth. Belonging to this class are the stream network patterns carved by groundwater erosion \cite{Deva2012,petroff}. 

Observations of such networks, in which groundwater flow being confined to shallow horizontal layers is  well-approximated by two-dimensional dynamics, has led to renewed interest in Loewner's equation governing the evolution of slits in the complex half-plane e.g. \cite{gubiec,Cohen,Deva}. Here the boundary formed by the growing slits is the interface, with an individual slit representing a stream. Slit-slit interaction  via the surrounding Laplacian field influences their growth and, in turn, the pattern formed by the network as a whole.

Mathematically, Loewner's equation is an ordinary differential equation (ODE) which governs the conformal map from a  domain characterised by slits to some canonical domain, usually a half-space or the  exterior of the unit disk. It links a real time-dependent function--the driving function--to a growing, continuous curve or slit in the complex domain. Since its introduction in the 1920s it has proved a useful tool in classical complex function theory--see \cite{abate} for a historical account. When the forcing function is a stochastic Brownian motion, the curve exhibits many special scaling properties which have been of considerable interest to mathematicians and physicists e.g. \cite{gruzberg,CARDY200581,lawler,bauer}.

This work considers the geodesic evolution of a slit in the upper half of the complex plane.
By `geodesic' it is meant that an individual slit follows a direction determined by the local Laplacian field at its tip  such that local symmetry is preserved or, equivalently, maximises the flux entering the tip \cite{selander,Carleson2002,Deva}. This requirement enables the forcing function  to be found  and ultimately the path of the growing slit. 

The growth of geodesic slits governed by Loewner's equation has been studied in various geometries including the half-plane, the whole plane minus, say, the positive real axis, and channels e.g. \cite{selander,Carleson2002,gubiec,duran2}. In all these examples the slits grow from a given interface along which the diffusive quantity $\phi$ is fixed e.g. the growth of slits from the real axis into the upper half-plane, with $\phi=0$ along the entire real axis and on the growing slit itself. The present work generalises this to slits growing in the upper half-plane from the real axis boundary where $\phi$ is piecewise constant. An immediate consequence of this boundary condition is that a single slit does not grow in a straight line parallel to the imaginary axis as it would do in the standard Loewner dynamics. Instead, it follows a curved path which ultimately may either  grow toward $y\to\infty$ for large time, or curve back toward the real axis terminating in finite time.  

Physically, the varying boundary condition for $\phi$ along the real axis may be thought of as a non-uniform source of $\phi$ along the domain boundary. The propagation of thin fingers in the presence of varying boundary conditions  has application in, for example, the evolution of groundwater-fed streams where localised sources and sinks (e.g. lakes) along the boundary influence the growth of nearby streams.

\section{Background}
\label{sec:2}
The Loewner differential equation
\begin{equation}
\frac{dg_t}{dt}=\frac{2}{g_t-a(t)},
\label{loewner}
\end{equation}
where $a(t)$ is a real-valued function, encodes the path of a slit in the upper half of the $z=x+iy$ plane growing from the starting point on the real axis $z=a(0)$. The function $w=g_t(z)$ maps the slit $z$-plane to the entire upper half of the $w$-plane with the property $g_t(z)=z+2t/z+O(1/z^{2})$ as $z\to\infty$. The tip of the slit, $z=\gamma(t)$, maps to $w=g_t(\gamma)=a(t)$ on the real $w$-axis.
Given the forcing function $a(t)$, and initial condition $g_0(z)=z$, the slit evolution in the $z$-plane can be calculated analytically in some cases, or numerically
 e.g. \cite{Kager2004,lind,Kennedy}.

The connection between Loewner dynamics and Laplacian growth is provided by the realisation that since $g_t(z)$ is a conformal map, $\phi(x,y)={\rm Im}(g_t)$ satisfies Laplace's equation
$\Delta \phi=0$ in the slit $z$-plane with properties $\phi\to y$ as $y\to \infty$ and $\phi=0$ both on the slit and the real axis. In order to complete the description, a rule for the direction of growth of the slit needs to be specified. This is equivalent to determining the forcing function in Loewner's equation (\ref{loewner}).

An important class of problems  governed by (\ref{loewner}) and its generalisation to $N$ slits (e.g. \cite{Kager2004}), considers  slits that grow along flow lines of the Laplacian field $\phi$ \cite{selander,Carleson2002,Deva}. That is, along  streamlines  $\psi(x,y)=constant$, where $\psi={\rm Re}(g_t(z))$ is the harmonic conjugate of $\phi$. Equivalently, the slit follows a path that maximises the flux entering its  tip \cite{Deva}. In these so-called geodesic growth problems, the form of the driving function is determined by the map $g_t(z)$, with exact solutions known for the symmetric growth of two and three slits e.g. \cite{selander,duran1,Deva}. Geometric properties of growing geodesic slits have successfully explained  features of Laplacian growth e.g. the properties of stream bifurcations in  groundwater-fed drainage systems \cite{Cohen,yi,Deva}.

Slit growth in other geometries may also be tackled using the Loewner formulation. For example, the geodesic growth of slits in semi-infinite strips with reflecting boundary conditions on the sides of the strip follows from deriving and solving a modified Loewner equation \cite{gubiec,duran2}. In the case when $\phi=0$ on all boundaries of a simply connected domain, geodesic paths are simply found by mapping from a half-plane to the slit domain of interest, with the paths being images under the conformal map. In this way, for example, the paths taken by  two slits emanating from the tip of a semi-infinite needle are obtained \cite{Deva,McD2020}. 

Geodesic evolution of a slit $\Gamma_t$ in the upper half of the $z$-plane $H$, governed by the following boundary value problem is considered:
\begin{align}
\Delta \phi&=0, \quad z\in H\backslash\Gamma_t, \nonumber \\
\phi&=h_0, \quad z\in\Gamma_t,\nonumber\\
\phi&=h(x),\quad z=x+0i,\nonumber \\
\phi&\to y, \quad {\rm as} \quad y\to\infty,
\label{bvp}
\end{align}
where $h_0=h(a(0))$ is a constant set by the starting location of the slit. There is a singularity in the gradient of $\phi$ at the tip of $\Gamma_t$ which causes it to grow \cite{gubiec}. While the  growth speed is unimportant for the problems considered here, its direction is given by the geodesic assumption. The problem (\ref{bvp}) is supplement by the usual initial condition $\Gamma_{t=0}\equiv 0$; that is, the slit begins growing into an empty half-plane.  

The essential difference in the  growth problem (\ref{bvp}) compared to that leading to the `standard' Loewner evolution is that the bottom boundary condition $\phi=h(x)$ replaces the more usual $\phi=0$. In \textsection\ref{sec:3} and \textsection\ref{traj} the choice 
\begin{align}
 h(x) = \begin{cases} c & |x|\le 1  \\
0 &  |x|>1, \end{cases}
\label{piecephi}
\end{align}
where $c$ is constant, is made. Note that  the boundary condition (\ref{piecephi}) implies that the slit path will in general follow a curved path except for the special case $a(0)=0$ where symmetry gives a  vertically growing slit.

Two approaches are used here to compute geodesic paths governed by (\ref{bvp}) and (\ref{piecephi}): (i) derivation and solution of a modified Loewner equation; (ii) using the standard Loewner equation (\ref{loewner}), together with a coupled set of ODEs which determine the forcing function $a(t)$. 

\section{Determining the path of the slit}
\label{sec:3}

\subsection{Derivation of a modified Loewner equation}{\label{sec:3a}}

\begin{figure}
\centering

  \includegraphics[width=0.8\textwidth]{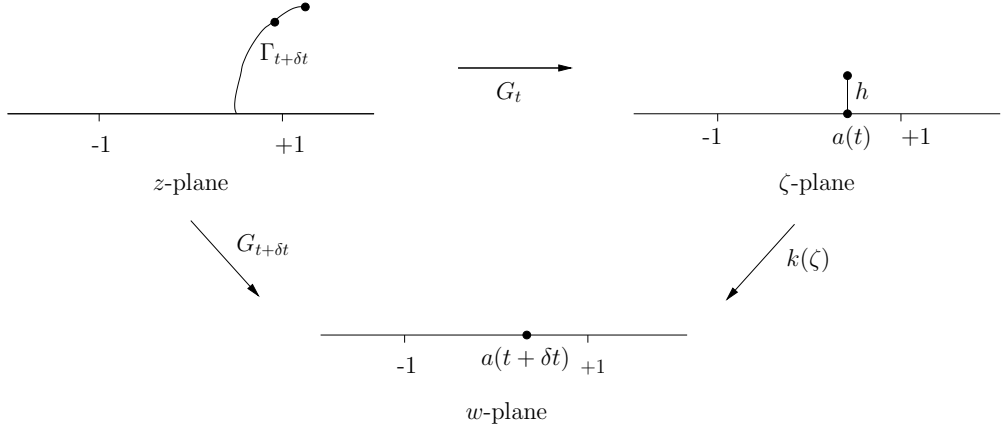}

\caption{Sequence of mappings used in deriving the modified Loewner equation. The points $\pm 1$ are invariant.}
\label{fig1}       
\end{figure}

Consider the sequence of maps shown in figure \ref{fig1}. The filled circles on the slit $\Gamma_{t+\delta t}$ in the $z$-plane indicate the position of the slit tip $\gamma_t$  at time $t$ and position $\gamma_{t+\delta t}$ a short time later $t+\delta t$. $\Gamma_{t+\delta t}$ is mapped to the upper half of the $w$-plane by $G_{t+\delta t}$, with its tip mapped to $w=a(t+\delta t)$. The same slit is mapped by $G_{t}$ to the $\zeta$-plane resulting in a small vertical increment of length $h$ which is the image under $G_t$ of the portion of the slit from $\gamma_t$ to $\gamma_{t+\delta t}$. Subsequently, $w=k(\zeta)$ maps the slit $\zeta$-plane to the upper half of the $w$-plane with  the property $k(\pm 1)=\pm 1$. It has the form
\begin{equation}
k(\zeta)=\frac{2\sqrt{(\zeta-a)^2+h^2}-\sqrt{(1-a)^2+h^2}-\sqrt{(1+a)^2+h^2}}{\sqrt{(1-a)^2+h^2}-\sqrt{(1+a)^2+h^2}}.
\label{rootmap}
\end{equation}
The appearance of the square root in (\ref{rootmap}) comes from the Schwarz-Christoffel map and its occurence  is standard in deriving Loewner's equation e.g. \cite{selander,gubiec}. Branches of the square roots are chosen such that $k(\pm 1)=\pm 1$, and $k(\zeta)\to \zeta$ for $\zeta\to\infty$, so that in the limit $h\to 0$  (\ref{rootmap}) becomes, retaining terms up to $h^2$,
\begin{equation}
k(\zeta)=\zeta+\frac{h^2(1-\zeta^2)}{2(1-a^2)(\zeta-a).}
\label{limitrootmap}
\end{equation}

Employing standard arguments used in deriving the Loewner equation and its variants e.g. \cite{gubiec}, 
the modified Loewner equation is obtained by considering the limit $\delta t\to 0$ of
\begin{equation}
\frac{dG_t}{dt}=\lim_{\delta t\to 0}\frac{k(G_t)-G_t}{\delta t}.
\label{limit}
\end{equation}
Using (\ref{limitrootmap}) in  (\ref{limit}) gives
\begin{equation}
\frac{dG_t}{dt}=\frac{1-G_t^2}{(G_t-a)(1-a^2)}=\frac{1}{\varphi (G_t)},
\label{loewnermod}
\end{equation}
where the assumption $h^2=2\delta t$ has been made.

To find $a(t)$, the procedure of Devauchelle {\it et al.} \cite{Deva} is used. This is based on the realisation that the small slit  in the $\zeta$-plane of figure \ref{fig1} is not precisely vertical, but rather it is curved along local streamlines and has a relatively small displacement of order $\delta t$ in the real direction. With this in mind, let $G_t(\gamma_{t+\delta t})=a_t+\beta\delta t+\alpha \sqrt{\delta t}$, where $\beta$ is a real and $\alpha$ is imaginary. Note the  $\sqrt{\delta t}$ reflects diffusive growth of the slits in the imaginary direction \cite{Deva},  and the choice $\beta\delta t$ gives a parabolic arc in the $\zeta$-plane for the image of the slit segment from $\gamma_t$ to $\gamma_{t+\delta t}$.  The curvature of the parabolic arc at $a(t)+i0$ is $\kappa_p=2\beta/|\alpha|^2$. The parabolic arc is consistent with the local curvature of the streamlines near the real axis which the slits initially follow. Devauchelle {\it et al}. \cite{Deva} show $\alpha^2=-1/\varphi'(a)$ and   $a(t)$ satisfies the ODE 
\begin{equation}
\frac{da}{dt}=2\beta-\frac{\varphi''(a)}{2{\varphi}'(a)^2}.
\label{adot}
\end{equation}
The  length of the arc displacement in the imaginary direction, $|\alpha| \sqrt{\delta t}$, is inconsistent with the choice made in deriving (\ref{loewnermod}) namely that  $h=\sqrt{2\delta t}$ since equating them gives $\alpha^2=-2$. But (\ref{loewnermod}) gives directly $\varphi'(a)=1$ and so, according to \cite{Deva} $\alpha^2=-1/\varphi'(a)=-1\ne -2$. The apparent contradiction is resolved by scaling $\varphi$ by $\lambda=\sqrt{2}$ and time by $\lambda^2=2$. Loewner's equation (\ref{loewnermod}) is invariant under this transformation (see also e.g. \cite{Kager2004}) and (\ref{adot}) becomes
\begin{align}
\frac{da}{dt^*}&=2\sqrt{2}\beta-\frac{{\varphi^*}''(a)}{2{\varphi^*}'(a)^2},\nonumber\\
&=2\sqrt{2}\beta-\frac{2a}{1-a^2},
\label{adot2}
\end{align}
where the superscript $*$ indicates the new scaled variables.
As before $\alpha^2=-1/{\varphi^*}'(a)=-1$, which is now consistent with the new timescale scaling $h=\sqrt{\delta t^*}=|\alpha|\sqrt{\delta t^*}$. Note the only difference in (\ref{adot2}) compared to (\ref{adot}) is the presence of the $\sqrt{2}$ factor multiplying the $\beta$-term. Henceforth the superscripts $*$ are dropped. Further evidence for the presence of the $\sqrt{2}$ factor is presented in \textsection\ref{traj} by comparing results for slit paths with those computed using a numerical method independent of Loewner's equation. 

Assuming $|a(t)|<1$ a further time transformation  is made to eliminate the factor $(1-a^2)$ in the denominators on the right-hand side of (\ref{loewnermod}) and (\ref{adot2}) giving
\begin{align}
\frac{dG_t}{dt}&=\frac{1-G_t^2}{G_t-a(t)},\label{gscaled}\\
\frac{da}{dt}&=2(1-a^2)\sqrt{2}\beta-2a.
\label{final2}
\end{align}
The case $|a(t)|<1$ is noted below. It remains to find an expression for $\beta$. In the standard geodesic Loewner problem the streamlines in the $\zeta$-plane are vertical, and so $\beta=0$. This is not the case here, and the following derives an expression for $\beta$ for geodesic paths for the variable boundary condition (\ref{piecephi}).

In the mapped $w=u+iv$ plane the streamlines are the real part of the analytic function 
\begin{equation}
F(w)=Kw+\frac{c}{\pi}\log\left [ \frac {w-1}{w+1}\right ],
\label{complexpot}
\end{equation} 
where $K=K(t)$.
Observe that (\ref{gscaled}) implies that $dG_t/dt\sim -G_t$ as $G_t\to\infty$. Together with the initial condition $G_0(z)=z$ this implies $w=G_t(z)=e^{-t}z$ as $z\to\infty$ and so $K=e^t$.

The streamline $\psi(u,v)={\rm Re}(F)=constant $ passing through $w=a+i0$ has curvature (noting $\partial\psi/\partial v=0$ at $(a,0)$) 
\begin{align}
\kappa&=-\left. \frac{\frac{\partial^2 \psi}{\partial v^2}}{\frac{\partial \psi}{\partial u}}\right|_{(a,0)} ,\nonumber \\
      &=\frac{-4ace^{-t}}{\pi(1-a^2)^2+2ce^{-t}(a^2-1)}.
\label{curvature}
\end{align}
Since $2\beta/|\alpha|^2=2\beta=\kappa_p$, equating $\kappa_p$ with $\kappa$ (\ref{curvature}) gives
\begin{equation}
2\beta=\frac{-4ace^{-t}}{\pi(1-a^2)^2-2ce^{-t}(1-a^2)}.
\label{twobeta}
\end{equation}
Substituting (\ref{twobeta}) into (\ref{final2}) yields finally an ODE for the driving function $a(t)$
\begin{equation}
\frac{da}{dt}=-2a+\frac{2\sqrt{2} a ce^{-t}}{c e^{-t}+\frac{\pi}{2} (a^2-1)},
\label{adotfinal}
\end{equation}
which when solved together with (\ref{gscaled}) gives the path of the slit.

If $|a(t)|>1$ then (\ref{gscaled}) becomes
\begin{equation}
\frac{dG_t}{dt}=\frac{G_t^2-1}{G_t-a(t)},
\label{loewnermodout}
\end{equation}
so $w=G_t(z)=e^{t}z$ as $z\to\infty$, and hence the ODE for the driving function is
\begin{equation}
\frac{da}{dt}=2a-\frac{2\sqrt{2} a ce^{t}}{c e^{t}+\frac{\pi}{2} (a^2-1)}.
\label{adotfinalout}
\end{equation}

\subsection{Standard Loewner equation method}{\label{sec:3b}}

An alternative approach is to use the standard Loewner equation (\ref{loewner}). In this case $\varphi (g_t)=(g_t-a)/2$, giving $\varphi''(a)=0$ and (\ref{adot}) becomes $da/dt=2\beta$. Note that $\beta$ is different to that in \textsection\ref{sec:3}\ref{sec:3a}, as is the timescale $t$.

The points $z=\pm 1$ at which the boundary values of $\phi$ jump are no longer invariant under the map $g_t$. Letting $w^{\pm}(t)=g_t(\pm 1)$, the complex potential in the $w$-plane is
\begin{equation}
F(w)=Kw+\frac{c}{\pi}\log\left [ \frac {w-w^+}{w-w^-}\right ].
\label{complexpot2}
\end{equation} 
Here, the choice $K=1$ is made since this gives the required condition $\phi\to y$ as $y\to\infty$ associated with (\ref{loewner}).
As in \textsection\ref{sec:3}\ref{sec:3a}, the curvature of the streamline $\kappa$ through $a+i0$ is computed from (\ref{complexpot2})
\begin{equation}
\kappa=\frac{-c[2a(w^+-w^-)+{w^-}^2-{w^+}^2]}{\pi(a-w^-)^2(a-w^+)^2+c(w^+-w^-)(a-w^-)(a-w^+)}.
\label{curvgen}
\end{equation}
Note if $w^{\pm}=\pm 1$, (\ref{curvgen}) reduces to (\ref{curvature}) up to the factor $e^{-t}$.
Using the result $\alpha^2=-1/\varphi'(a)=-2$ \cite{Deva}, and $2\beta/|\alpha|^2=\kappa_p=\kappa$, gives, from (\ref{curvgen}) and (\ref{adot}),
\begin{equation}
\frac{da}{dt}=2\sqrt{2}\beta=\frac{-2c\sqrt{2}[2a(w^+-w^-)+{w^-}^2-{w^+}^2]}{\pi(a-w^-)^2(a-w^+)^2+c(w^+-w^-)(a-w^-)(a-w^+)}.
\label{adotfinal2}
\end{equation}
The $\sqrt{2}$ factor in (\ref{adotfinal2}) is again needed to make the scaling for $h$ in the derivation of (\ref{loewner}) and vertical scale $|\alpha|\sqrt{\delta t}$ in the derivation of (\ref{adot}) consistent.
Equation (\ref{adotfinal2}) along with the pair of evolution equations for $w^{\pm}(t)$ obtained from (\ref{loewner})
\begin{equation}
\frac{dw^{\pm}}{dt}=\frac{2}{w^{\pm}-a},
\label{wplusminus}
\end{equation}
represent a coupled system of three ODEs for $a, w^{\pm}$ which when solved subject to initial conditions $w^{\pm}(0)=\pm 1$, $a(0)=a_0$ determine the forcing $a(t)$  in Loewner's equation (\ref{loewner}). The system (\ref{loewner}), (\ref{adotfinal2}) and (\ref{wplusminus}) applies equally for  $|a(t)|<1$ and $|a(t)|>1$.

\section{Slit trajectories}
\label{traj}

\subsection{The case $c=-1$}

Both formulations derived in \textsection\ref{sec:3} are used to find the trajectories of a single slit with different starting locations along the real axis and $c=-1$. This is done using the numerical procedure described in Kager  {\it et al}. \cite{Kager2004}: for a given time $t$, the discretised version of (\ref{loewner}) or (\ref{loewnermod}) is integrated backwards in time from the condition $g_t=a(t)$, or $G_t=a(t)$, to obtain the value of $g_0=\gamma(t)$ or $G_0=\gamma(t)$ i.e. the location of the tip of the slit in the $z$-plane. This requires the driving function to be known in advance either by numerically solving the ODE for the driving function (\ref{adotfinal}), or the coupled set of ODEs for $a(t)$ and $w^{\pm}$, (\ref{adotfinal2}) and (\ref{wplusminus}).

\begin{figure}
\centering

  \includegraphics[width=.8\textwidth]{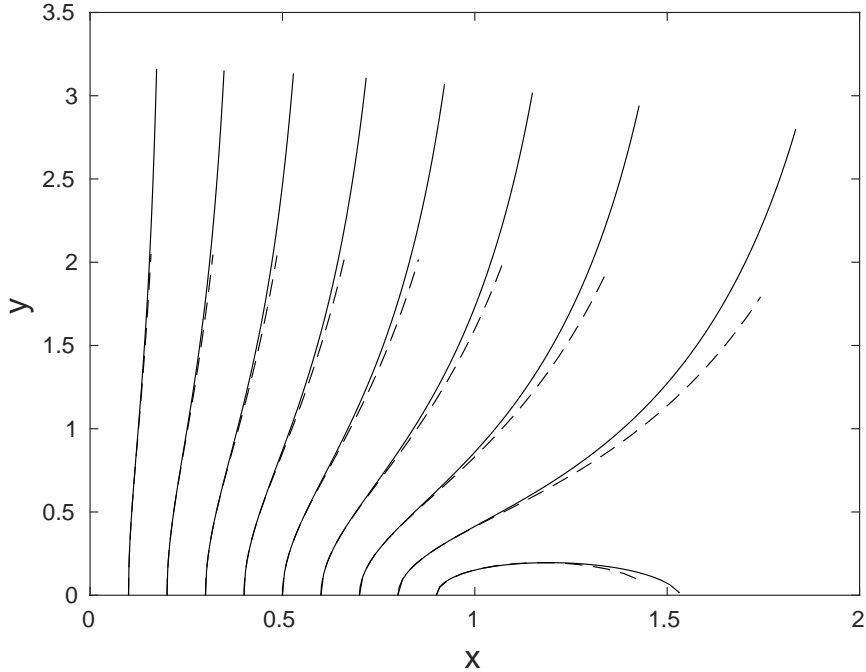}

\caption{Slit trajectories for $c=-1$ for a range of starting location on the real axis computed by solving Loewner's equation (\ref{loewnermod}) with forcing (\ref{adotfinal}) (solid line). Also shown are trajectories for the same selection of starting locations computed using the numerical method \cite{McD2020} (dashed lines). }
\label{comparison}       
\end{figure}

While the timescales are different, the paths of the slits obtained via the standard and modified Loewner equations (\ref{loewner}) and (\ref{loewnermod}) agree with each other to a high degree of accuracy, and in the figures that follow the slit paths obtained by either method coincide. Figure \ref{comparison} shows examples of paths (solid lines) taken by a single growing slit with different starting locations $0<a(0)<1$ along the positive real $z$-axis. Results in this paper for single slit evolution are only discussed for slits starting along the positive real axis, since those starting on the negative real axis are  obtained by symmetry. 

As a check, superimposed on this plot are the trajectories, shown as dashed lines, computed using the numerical procedure \cite{McD2020} which approximates the growing slits as an increasing number  of small straight line segments. In the experiments shown each segment has  length  0.05. The straight line segment approximation enables the slit  to be mapped to the upper half of the $w$-plane by the  Schwarz-Christoffel map. This is done numerically using Driscoll's SC Toolbox \cite{SCT}. In the $w$-plane, the solution (\ref{complexpot2}) is used to extend the solution along contours of $\psi$. The SC Toolbox maps the new point in the $w$-plane  back to the $z$-plane giving the next segment of the growing slit. Since the method \cite{McD2020} does not solve any form of Loewner's equation, nor an equation for the forcing function, it is   a useful independent check on the paths calculated using the Loewner equation based formulations of \textsection\ref{sec:3}. As shown in figure \ref{comparison} agreement between the numerical procedure \cite{McD2020} and that of the Loewner equation methods of the present work is reasonable. There is some expected loss of accuracy in the numerical procedure \cite{McD2020} when approximating a path by straight line segments as curvature increases.

\begin{figure}
\centering

  \includegraphics[width=.8\textwidth]{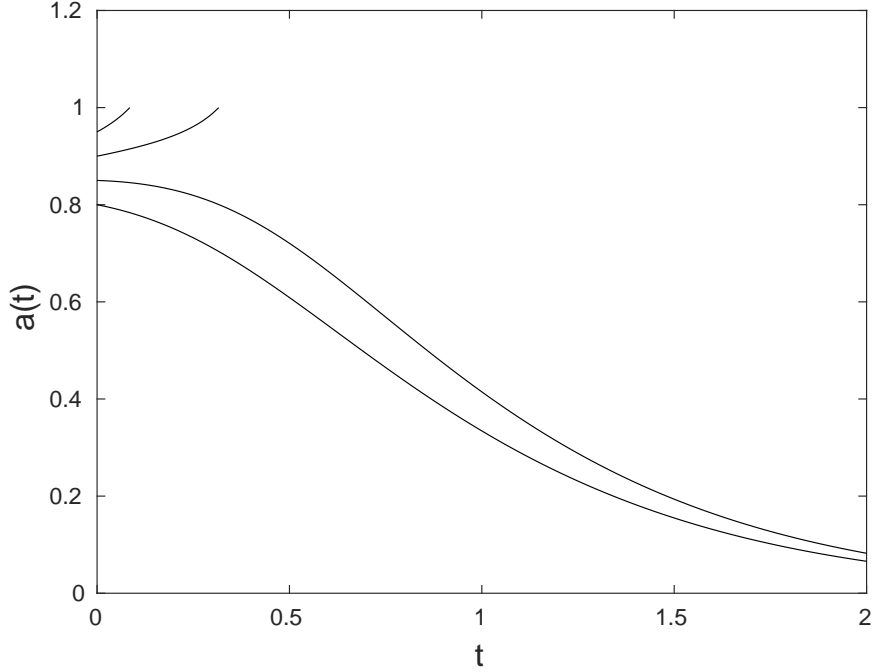}

\caption{Forcing function $a(t)$ as a function of time for $c=-1$, and $a(0)=$0.8, 0.85, 0.9 and 0.95. Numerical solutions for $a(0)=0.9$ and $a(0)=0.95$ are terminated at $t\approx0.316$ and $t\approx0.085$ when they reach $a(t)=1$.
 }
\label{forcetime}       
\end{figure}

The behaviour of the slit paths in figure \ref{comparison} depends on their starting location, with some paths heading toward $y\to\infty$ while curving toward increasing values of $x$. In contrast, paths starting near the jump in boundary value at $z=1$ turn over and collide with the real axis e.g. the path starting at $a(0)=0.9$. From   (\ref{adotfinal}), at $t=0$, $da/dt$  changes sign from negative to positive at $a(0)=\sqrt{1+2c(\sqrt{2}-1)/\pi}\approx 0.858$.   This suggests that near this value, $a(t)$ may either decrease or increase, and if it increases to unity then the solution breaks down owing to  the $1-a^2$ factor in (\ref{loewnermod}). This is confirmed by numerical solution of (\ref{adotfinal}): $a(t)$ decays  to zero as $t\to 0$ for $a(0)<0.885$ and increases for  $0.885<a(0)<1$, reaching unity in a finite time. Example plots of $a(t)$ as a function of time near this transition point are shown in figure \ref{forcetime} for the choices $a(0)=$0.8, 0.85, 0.9 and 0.95. For $a(0)=$0.8 and 0.85, $a(t)$ decreases, whereas for $a(0)=$0.9 and 0.95, $a(t)$ increases, reaching unity in finite time $t\approx0.316$ and $t\approx0.085$ respectively. While the solution for $a(t)$ can be continued beyond these times, the solution of the trajectory breaks down coinciding with its collision with the real axis. 

\begin{figure}
\centering
  \includegraphics[width=.8\textwidth]{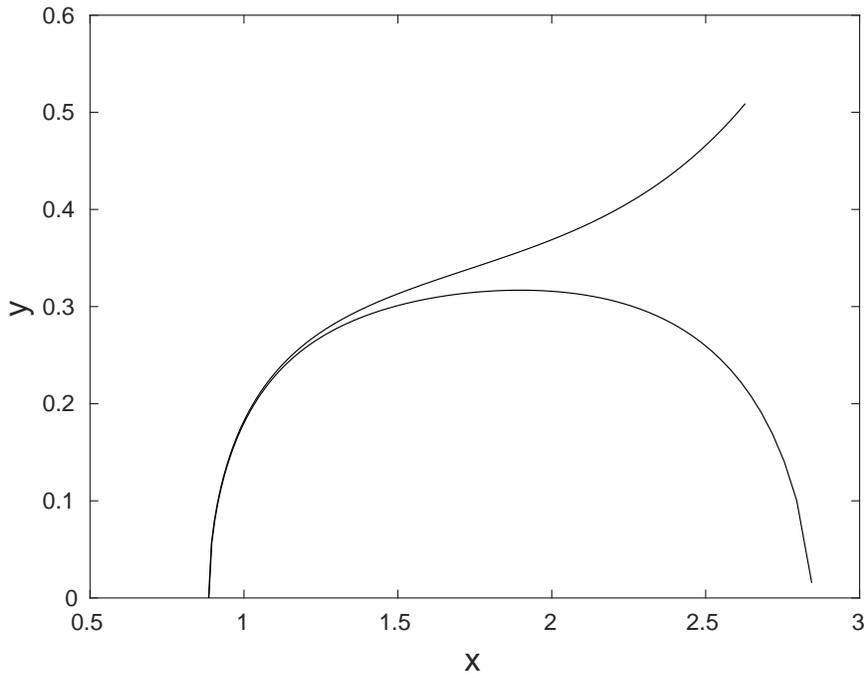}

\caption{Comparison of slit trajectories with $c=-1$ up to time $t=0.755$ starting at (i) $a(0)=0.885$ which eventually grows upwards, and (ii) $a(0)=0.886$ which eventually collides with the real $z$-axis.}
\label{figclose}       
\end{figure}

The change in nature of the slit paths according to an increasing or decreasing forcing function is shown in figure \ref{figclose} for trajectories starting at two closely located starting points $a(0)=0.885$ and $a(0)=0.886$: when $a\to 0$ then the path trajectory eventually grows to infinity in the positive imaginary direction. In contrast, if $a(t)$ increases then the path collides with the real $z$-axis in finite time, and this time coincides with the time at which $a(t)=1$. In this example this happens when $t\approx 0.755$. 

The trajectories shown in figure \ref{comparison} show that the slits, which carry value of $\phi=-1$ for this range of starting locations,  grow  toward regions where $\phi$ is greater. Depending on how close they are to the jump in $\phi$ at $z=1$ this may mean they curve back toward the real axis where $\phi=0$ for $|x|>1$, or continue to grow to infinity, where $\phi\to y$.

\begin{figure}
\centering

  \includegraphics[width=.8\textwidth]{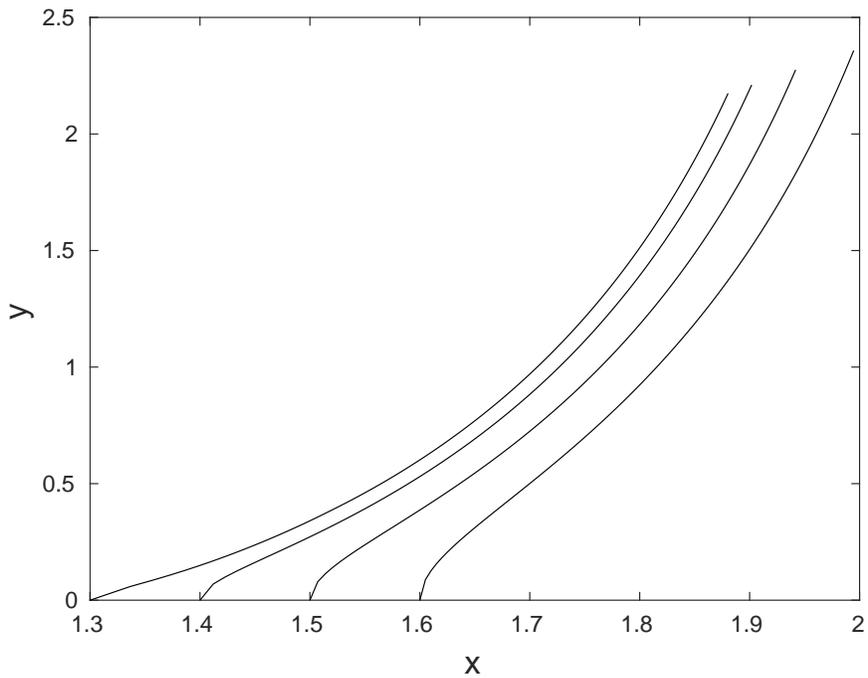}

\caption{Example slit trajectories for $c=-1$ and values of $a(0)>1$.}
\label{traj1}       
\end{figure}

Figure \ref{traj1} shows example trajectories with starting locations 1.3, 1.4, 1.5 and 1.6 on the real axis. All trajectories eventually grow upward toward $y\to\infty$ while moving toward more positive vales of $x$. This is expected since for these starting locations $\phi=0$ on the slits and they move away from the section $|x|<1$ on which $\phi=-1$. Numerical experiments show that if the starting point is in the range $1<a(0)< a_{crit}$, where $a_{crit}\approx 1.28$ none of the numerical procedures of \textsection\ref{sec:3} converge, suggesting that no trajectories leaving the real axis are possible. The reason for this non-existence is apparent from (\ref{complexpot}): at $t=0$ (i.e. when $z\equiv w$) the vertical velocity along the real $z$-axis is 
\begin{equation}
V=\frac{dF}{dz}=1+\frac{2c}{\pi(x^2-1)}.
\label{vertneg}
\end{equation}
For $c=-1$, (\ref{vertneg}) implies that $V<0$ for $1<x<\sqrt{1+2/\pi}$. Slits are unable to form when $V<0$ and so $a_{crit}=\sqrt{1+2/\pi}\approx 1.2793$, consistent with the numerical result that convergent numerical paths cannot be computed for  this range of starting locations.

\subsection{The case $c=+1$}

In this case the vertical velocity along the real axis at $t=0$ is from (\ref{vertneg})
\begin{equation}
V=1+\frac{2}{\pi(x^2-1)},
\label{vertplus}
\end{equation}
which is positive for $0\le x< a_{crit}$ and $x>1$, where $a_{crit}=\sqrt{1-2/\pi}\approx 0.603$. Thus Loewner slits are unable to grow from the interval
$a_{crit}\le a(0) \le 1$. 

\begin{figure}
\centering

  \includegraphics[width=.8\textwidth]{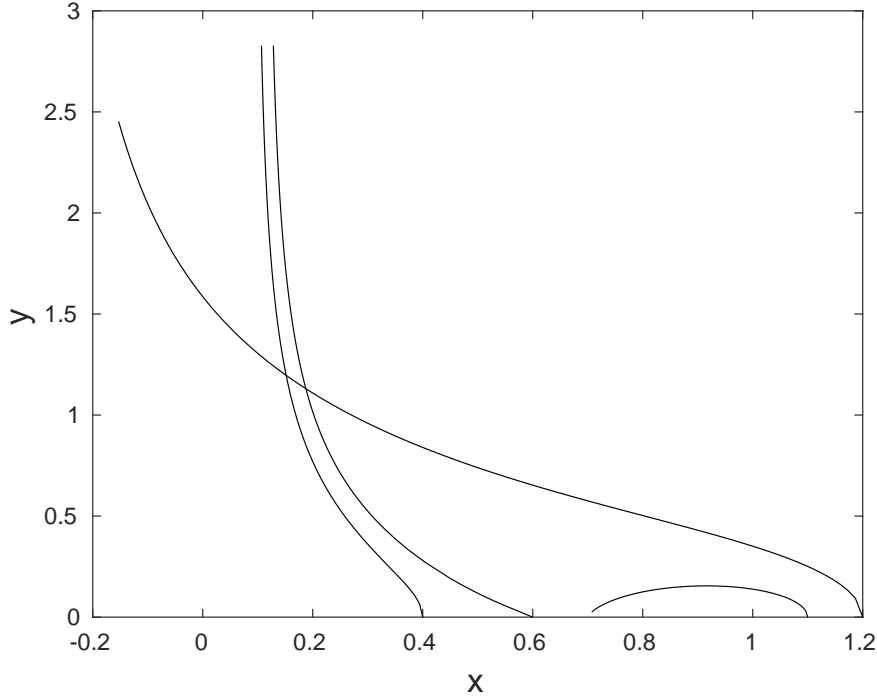}

\caption{Path  of slits for $c=+1$ and $a(0)=0.4$, $a(0)=\pm 0.6$,  and $a(0)=1.2$  over time $0<t<2$. Also shown is the path for $a(0)=1.1$ for $0<t<0.24$}

\label{cplustraj}       

\end{figure}

Figure \ref{cplustraj} shows trajectories of slits starting at  locations either side of the forbidden range $a_{crit}\le a(0) \le 1$. They all tend toward the negative $x$-direction as they grow upward, since this is the direction where the background $\phi$ field is larger. The slit starting at $a(0)=1.1$ turns over,  eventually colliding with the real axis at $t\approx 0.24$. As in the case $c=-1$, this behaviour is understandable from (\ref{adotfinalout}) which shows that at $t=0$,  $da/dt$ changes sign when $a(0)=\sqrt{1+2(\sqrt{2}-1)/\pi}\approx 1.12$. When $1<a(0)<a_c$, $a(t)$ decreases to unity in finite time, whereupon $da/dt$ becomes unbounded and the slit trajectory  collides with the real axis. Numerical solution of (\ref{adotfinalout}) gives $a_c\approx 1.11$.

\subsection{Pairs of slits}{\label{slitpairs}}

With the standard boundary condition $\phi=0$ on the real $z$-axis, the Loewner equation describing the geodesic growth of two symmetric slits in the half-plane admits an exact solution \cite{gubiec,Deva}. It is found that the slits curve away from each other and asymptote to straight lines with angle of $\pi/5$ between them. The asymptotic behaviour of the paths of the slits in this and other geometries has  application to patterns formed by groundwater-fed stream networks e.g. \cite{Deva2012,Cohen,yi}.
Similarly, the  growth of symmetric pairs of slits is considered here with the piecewise constant boundary condition (\ref{piecephi}) using a modification of the equations in \textsection\ref{sec:3}\ref{sec:3b}. If the slits grow from $z=\pm a(0)$, the Loewner equation governing the map $w=g_t(z)$ from the slit $z$-plane to the upper half of the $w$-plane is
\begin{equation}
\frac{dg_t}{dt}=\frac{2}{g_t-a(t)}+\frac{2}{g_t+a(t)}=\frac{4g_t}{g_t^2-a^2}=1/\varphi(g_t).
\label{2loewner}
\end{equation}
Equation (\ref{2loewner}) gives $\varphi'(a)=1/2$ and $\varphi''(a)=-1/2a $, and (\ref{adot}) gives
\begin{equation}
\frac{da}{dt}=\frac{1}{a}+2\beta_2,
\label{adotpair}
\end{equation}
 where $\beta_2$ is related to the curvature of the streamline at $w=a(t)$. As in \textsection\ref{sec:3}\ref{sec:3b} the streamlines in the $w$-pane are level curves of the real part of the complex potential (\ref{complexpot2}). The same arguments for re-scaling $\varphi$ and time apply as in \textsection\ref{sec:3}\ref{sec:3b} and (\ref{adotpair}) becomes
\begin{align}
\frac{da}{dt}&=\frac{1}{a}-\frac{2c\sqrt{2}[2a(w^+-w^-)+{w^-}^2-{w^+}^2]}{\pi(a-w^-)^2(a-w^+)^2+c(w^+-w^-)(a-w^-)(a-w^+)},\nonumber \\
             &=\frac{1}{a}-\frac{8\sqrt{2}cw^+a}{\pi(a^2-{w^+}^2)^2 +2cw^+(a^2-{w^+}^2)},
\label{adotpairfinal}
\end{align}
since $w^-=-w^+$ by symmetry. The evolution equation for $w^+(t)=g_t(1)$ follows from (\ref{2loewner})
\begin{equation}
\frac{dw^+}{dt}=\frac{4w^+}{{w^+}^2-a^2}.
\label{what}
\end{equation}
The coupled system (\ref{2loewner}), (\ref{adotpairfinal}) and (\ref{what}) is solved numerically to determine the trajectories of the slit pairs.

\begin{figure}
\centering

  \includegraphics[width=.8\textwidth]{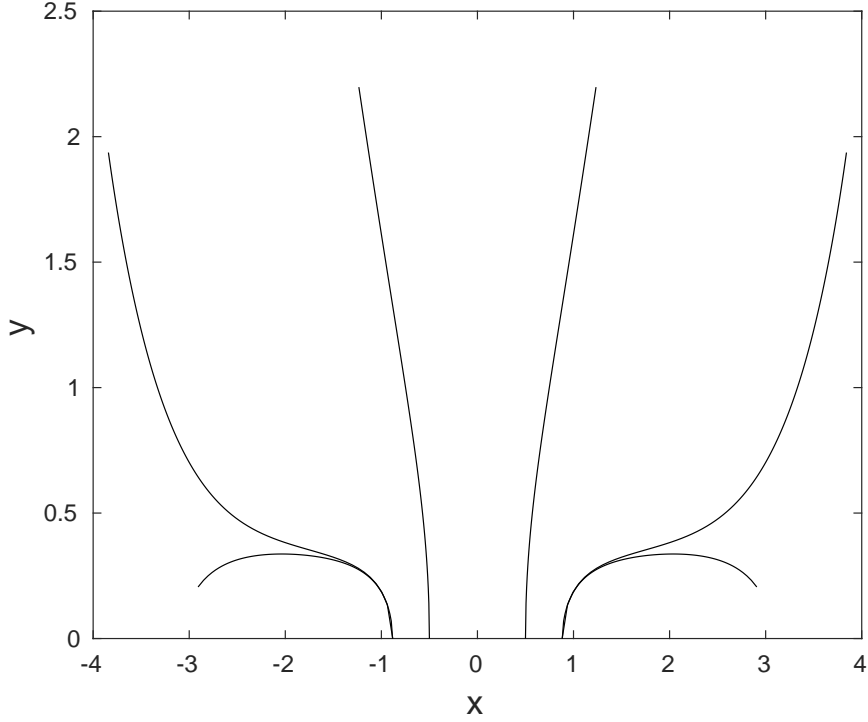}

\caption{Paths for symmetric pairs of slits for $c=-1$ and $a(0)=\pm 0.5$, $a(0)=\pm 0.882$ for $0\le t\le 1$, and $a(0)=\pm 0.883$ for $0\le t\le 0.125$.}

\label{figpairs}       

\end{figure}

Choosing $c=-1$, figure \ref{figpairs} shows pairs of slits for three different numerical experiments each having different initial starting locations of the slits: $a(0)=\pm 0.5$, $a(0)=\pm 0.882$, and $a(0)=\pm 0.883$. In each case the pairs initially grow upwards  while curving away from each other under the influence of both the non-uniform boundary condition on the real axis and slit-slit interaction. As demonstrated in figure \ref{figpairs}, there is a distinct change in behaviour of the paths when the initial starting location  $a(0)\approx 0.882$: the pair starting closer to $z=1$ eventually turns and heads back toward, and ultimately collides with, the real axis  at time $t\approx 0.125$. The numerical solution shows that at this time $a(t)-w^+\to 0$ and so, from (\ref{adotpairfinal}) and \ref{what}), both $da/dt$ and $dw^+/dt$ become unbounded.

\begin{figure}
\centering
\begin{subfigure}[h]{0.49\linewidth}
 \includegraphics[width=\linewidth]{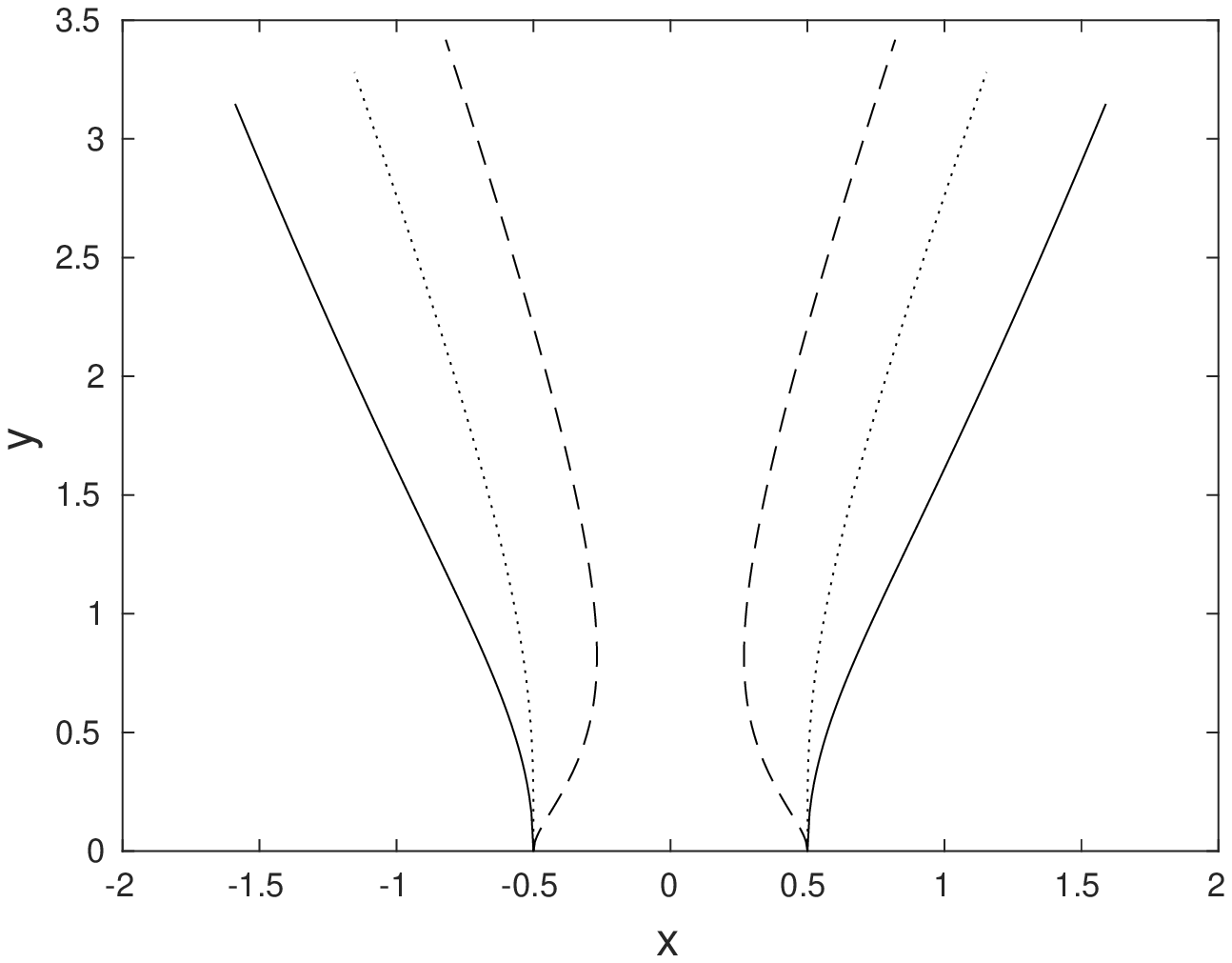}
\caption{}
\label{paircomparisona}
\end{subfigure}
\hfill
\begin{subfigure}[h]{0.49\linewidth}
 \includegraphics[width=\linewidth]{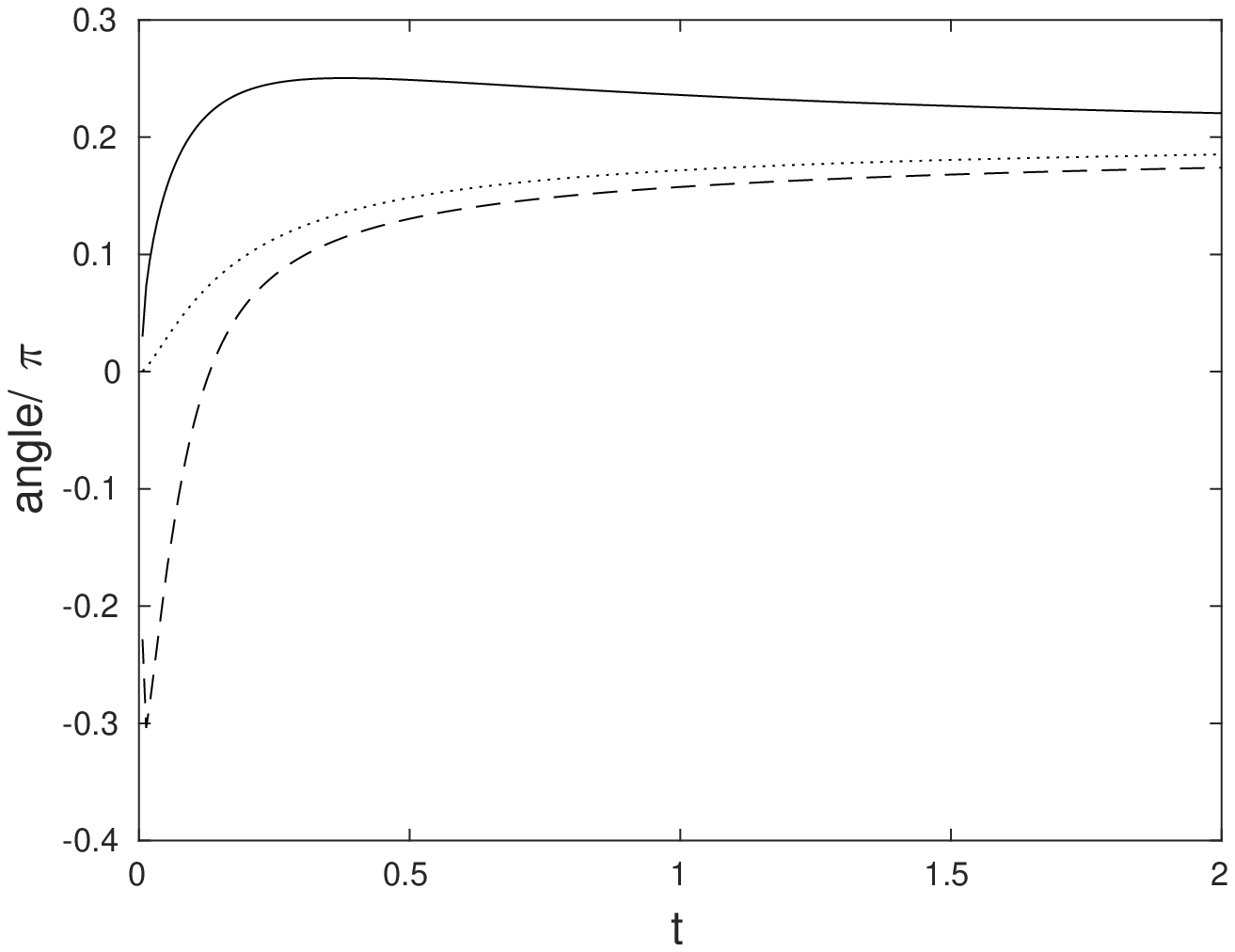}
\caption{}
\label{paircomparisonb}
\end{subfigure}
\caption{(a)  Paths of pairs of slits for  $a(0)=\pm 0.5$ and $c=-1$ (solid), $c=+1$ (dashed) and $c=0$ (dotted) for $0\le t\le 2$.  (b) Angle between trajectories as a function of time: $c=-1$ (solid), $c=+1$ (dashed) and $c=0$ (dotted). }
\label{paircomparison}
 \end{figure}

Figure \ref{paircomparisona} compares trajectories of symmetric pairs of slits for $c=\pm 1$, and for the standard Loewner evolution when $c=0$. While the non-uniform boundary condition influences the initial paths taken by the slits by either initially curving them toward or away from the symmetry axis, they  eventually tend to a constant angle of inclination with respect to each other independent of the value of $c$. The time evolution of the angle between the slits (as measured by the angle between the tangents at the slit tips) for each case is shown in figure \ref{paircomparisonb}.
It is well-known that for $c=0$ the angle approaches $\pi/5$ as $t\to\infty$ e.g. \cite{gubiec,Deva}. The pairs with the varying boundary conditions $c=\pm 1$ also  approaches the same angle; once the paths are sufficiently distant from the real axis, the boundary condition on the real axis becomes less influential and the pair behaves  like the standard $c=0$ case in approaching the $\pi/5$ angle between slits.

\section{Dipole on the $z$-axis}
\label{sec:4}

The effect of a dipole centred on the real $z$-axis on the geodesic slit growth is considered. This is a limiting case of the piecewise constant boundary condition considered in \textsection\ref{sec:3} when the length of interval on which $\phi=c\ne 0$ shrinks to zero and $c\to\infty$. As in \textsection\ref{sec:3}\ref{sec:3a} an ODE for the forcing function in (\ref{loewner}) is found by considering the curvature of the streamlines in the $w$-plane.  Let the dipole be placed at $z=0$ and have complex potential $q/z$ where $q\in {\rm Re}$.
As usual, $w=g_t(z)$ maps the slit $z$-plane to the upper half of the $w$-plane with the dipole mapping to $w=s$ on the real $w$-axis. The complex potential in the $w$-plane is then 
\begin{equation}
F=w+\frac{q}{w-s}.
\label{complexpotdip}
\end{equation} 
 Note that $\phi={\rm Im}F=0$ on the real $w$-axis. The geodesic slit path is therefore determined by (\ref{loewner}) with  $a(t)$ given by solution of
\begin{equation}
\frac{da}{dt}=2\sqrt{2}\beta,
\label{adotdip}
\end{equation}
where, as before, $\kappa=2\beta/|\alpha|^2$ is  the curvature of the streamline through $a+i0$ in the $w$-plane. In writing (\ref{adotdip}) it has been assumed, as in \textsection\ref{sec:3} that $\varphi(g_t)$ and $t$ have been rescaled giving the factor of $\sqrt{2}$ on the right-hand side of (\ref{adotdip}). 

Substituting $s(t)=g_t(0)$ into Loewner's equation (\ref{loewner}) gives an ODE for the dipole location:
\begin{equation}
\frac{ds}{dt}=\frac{2}{s-a}.
\label{cdotdip}
\end{equation}

In the $w$-plane the streamfunction is, from (\ref{complexpotdip}),
\begin{equation}
\psi(u,v)={\rm Re}(F)=u+\frac{q(u-s)}{(u-s)^2+v^2},
\label{streamdip}
\end{equation}
from which the curvature of the streamline through $a+i0$ is
\begin{equation}
\kappa=\frac{2q}{(a-s)\left[(a-s)^2-q\right]}.
\label{curvdip}
\end{equation}
Substituting (\ref{curvdip}) into (\ref{adotdip}), after noting $|\alpha|^2=1/\varphi'(a)=2$ and $\beta=\kappa$, gives
\begin{equation}
\frac{da}{dt}=\frac{4\sqrt{2}q}{(a-s)[(a-s)^2-q]}.
\label{adotdipfinal}
\end{equation}
The coupled system of equations (\ref{loewner}), (\ref{cdotdip}) and ({\ref{adotdipfinal}) determines the evolution of the slit, and is solved numerically as in \textsection\ref{traj}.

For the choice $q=-1$, the streamlines in the $w$-plane given by (\ref{streamdip}) are  those owing to a dipole in a uniform stream representing flow past a circular cylinder of unit radius centred at $w=s$. Streamlines to the right of $w=s$, curve left toward the dipole axis as they progress upward in the $w$-plane so it is expected that the slit will also curve in this direction as it grows. This is demonstrated in figure (\ref{dipneg}) showing slit paths for various starting positions. Sufficiently close to the dipole the slit curves back toward the real axis e.g. the slit path starting at $a(0)=1.3$ in figure \ref{dipneg}. This trajectory terminates in a numerically-determined finite time $t\approx 1.402$ before reaching the real axis.

\begin{figure}
\centering

  \includegraphics[width=.8\textwidth]{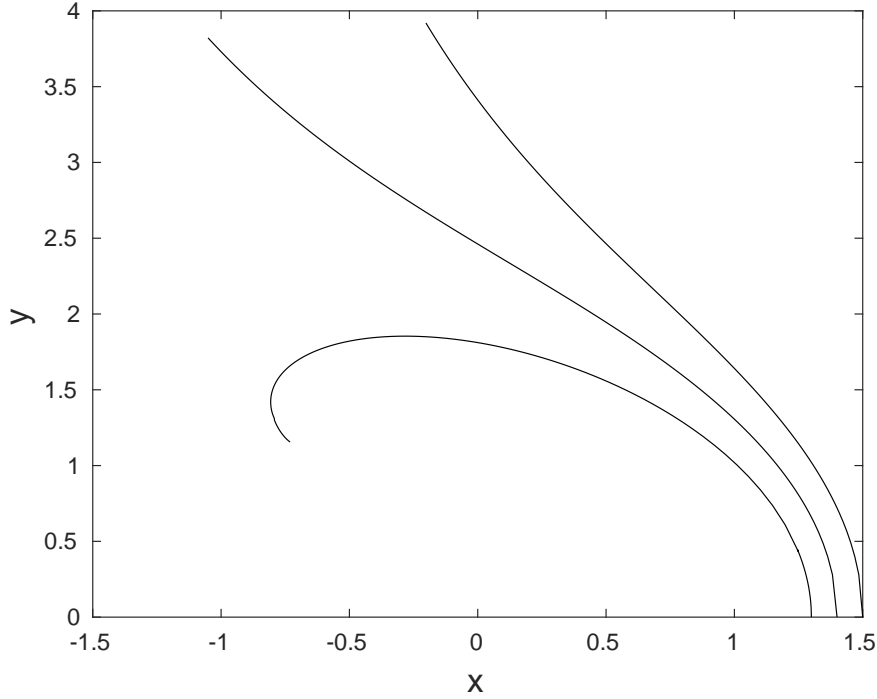}

\caption{Slit paths for $0\le t\le 4$ starting at 1.4 and 1.5 in the presence of a dipole of strength $q=-1$ at $z=0$. Also shown is the trajectory starting at 1.3 over the time $0\le t\le t_c\approx 1.402$}

\label{dipneg}       

\end{figure}

Insight into the finite lifetime of some trajectories is gained by combining (\ref{cdotdip}) and (\ref{adotdipfinal}) to give a single ODE for $b(t)=a(t)-s(t)$:
\begin{equation}
\frac{db}{dt}=\frac{(4\sqrt{2}-2)q+2b^2}{b(b^2-q)},
\label{beq}
\end{equation}
with solution
\begin{equation}
-2\sqrt{2}q\log|b^2+(2\sqrt{2}-1)q|+b^2=4t+A,
\label{beqsol}
\end{equation}
where $A$ is a constant determined by the initial condition $b(0)=a(0)-s(0)$.  The  transition between paths growing upward  or curling back toward the real axis, as shown in figure \ref{dipneg}, is able to be deduced from (\ref{beqsol}).
First, since $s(0)=0$ and $a(0)>0$, it follows $b(0)>0$ and  from (\ref{beq}) with $q=-1$, $b(t)$  decreases or increases if it starts either to the left or right respectively of the critical point $b(0)=\sqrt{2\sqrt{2}-1}=b_c\approx 1.35$. This value of $b_c$ is consistent with  figure \ref{dipneg} showing a change in the behaviour of the  paths with initial conditions $b(0)=a(0)-s(0)=1.3$ and $1.4$. If $b(t)$ decreases, as it does for $b(0)<b_{c}$, according to (\ref{beqsol}) it vanishes in finite time $t_c$, whereupon the solution breaks down since the denominator on the right-hand side of either (\ref{cdotdip}) and ({\ref{adotdipfinal}) vanishes. This is the case in the example shown in figure \ref{dipneg} with $b(0)=1.3$. Equation (\ref{beqsol}) gives the breakdown time $t_c$  explicitly:
\begin{equation}
-2\sqrt{2}q\log|b^2+(2\sqrt{2}-1)q|+b^2=4t_c-2\sqrt{2}q\log|a(0)^2+(2\sqrt{2}-1)q|+a(0)^2.
\label{crit}
\end{equation}
For example, if $a(0)=1.3$ (\ref{crit}) gives $t_c\approx 1.402$ consistent with the numerical determination of the breakdown time for the trajectory shown in figure \ref{dipneg}.

For the case $q=+1$,  at $t=0$ the velocity $V$ in the imaginary direction on the real $z$-axis owing to a dipole at the origin is $V=1-q/x^2$. Thus if $|a(0)|< 1$, $V<0$ and slits are unable to grow. This is confirmed by numerical experiment, which fails to find convergent solutions to (\ref{loewner}) and (\ref{adotdipfinal}) if $|a(0)|< 1$. For $|x|>1$, $V>0$ and the streamlines of the dipole field grow upward and away from the dipole toward increasing values of $x$. Thus it is expected that slits with starting location $|a(0)|> 1$ will curve away from the dipole at the origin as they grow upwards. This is the case, and slit trajectories (not shown) are qualitatively similar to those shown in figure \ref{traj1}.

\section{Conclusion}
\label{sec:5}

The present method extends to  boundary conditions where $\phi$ is a general piecewise-constant function on the real axis:
\begin{equation}
\phi=c_i,\quad x_i<x<x_{i+1},\quad i=1,\cdots,N-1,
\label{bgen}
\end{equation}
and $\phi=c_0$, $x<x_1$ and $\phi=c_N$, $x>x_N$. To proceed, the slit $z$-plane is mapped to the upper half of the $w$-plane and the curvature of the streamline through $w=a(t)$, where the streamfunction is the real part of the complex potential
\begin{equation}
F(w)=\frac{1}{\pi}\sum_{i=1}^N (c_{i-1}-c_i)\log(w-b_i),
\label{piecepot}
\end{equation}
and $b_i=g_t(x_i)$. A system of $N+1$ coupled ODEs  for $b_i(t)$, $i=1,\cdots,N$ and $a(t)$ is obtained by substituting into (\ref{loewner}). Their solution gives $a(t)$ which, in turn, is used to compute the slit path using the Loewner equation (\ref{loewner}). The alternative method of deriving a modified Loewner equation  does not seem straightforward for these more complicated boundary conditions since this would require a more involved sequence of transformations than that illustrated in  figure \ref{fig1}.

The piecewise constant distribution of $\phi$, and the dipole on the real axis are special cases of the general condition $\phi=h(x)$ on $y=0$. In the mapped $w$-plane, $\phi={\tilde h}(u)=h({\rm Re}(g_t(z)))$ along the real axis and the solution to Laplace's equation in the half-plane, such that $\phi\to v$ as $v\to\infty$, is
\begin{equation}
\phi=v+\frac{v}{\pi}\int_{-\infty}^{\infty}\frac{{\tilde h}(\zeta)}{(u-\zeta)^2+v^2}\, d\zeta,
\label{phigen}
\end{equation}
with harmonic conjugate
\begin{equation}
\psi=u-\frac{1}{\pi}\int_{-\infty}^{\infty}\frac{{\tilde h}(\zeta)(u-\zeta)}{(u-\zeta)^2+v^2}\, d\zeta.
\label{psigen}
\end{equation}
For example, choosing ${\tilde h}(\zeta)=Q\delta(\zeta-s)$ recovers $\psi$ and $\phi$ for the dipole (\ref{complexpotdip}) with $Q/\pi=-q$. The general problem  can be tackled numerically by discretising $h(x)$ to obtain the equivalent piecewise constant boundary value problem (\ref{bgen}) and (\ref{piecepot}). A further application of the second approach where the standard Lowener equation is solved with a suitably modified driving function, would be to consider channel geometries with periodic or reflecting boundary conditions \cite{gubiec}.

An interesting inverse problem is to ask the question: `what $h(x)$ is required to reproduce a given geodesic path?' In the standard Loewner problem, Kennedy \cite{Kennedy} addresses numerically the similar task of finding the forcing $a(t)$  given the path of a slit (see also Tsai \cite{tsai} for paths on lattices).   For example, in the standard Loewner evolution it is known that $a(t)=t$ is exactly solvable and gives a path which  asymptotes to $z=x+i\pi$, $x\to+\infty$  \cite{Kager2004}. Is there a $h(x)$ that implies ${\dot a}=1$ and therefore reproduces this path? Addressing such questions opens up the possibility of controlling the paths taken by slits through choice of the boundary condition on the real axis. 

As remarked in  \textsection\ref{intro}, non-uniform $\phi$ along the real axis represents  sources and sinks of the diffusive phase along the boundary of the domain. In application to the growth of stream networks formed by groundwater erosion, the results  indicate the possibility of regions for which streams do not form, or streams which either grow to large distance from their origin, or  turn back toward the boundary. It is significant that streams which do grow to large distances asymptote toward paths with angle of $\pi/5$ between them. Since the results here extend  to other geometries by conformal mapping, this  implies a bifurcation angle of $2\pi/5$ for stream pairs bifurcating from another `semi-infinite' stream. The $2\pi/5$ angle is known to be special in the geometry of drainage networks formed by seepage erosion \cite{Deva2012,petroff}, and that in standard Loewner dynamics it is a stable fixed point in bifurcating streams \cite{selander,Deva}.

\bibliographystyle{unsrt}


\end{document}